\documentclass[aip,jmp,showpacs,amsmath,a4paper,superscriptaddress,floatfix,notitlepage,12pt]{revtex4-1}

\usepackage{amssymb}
\usepackage{amsmath}
\usepackage{graphicx}
\usepackage{amsfonts}
\usepackage{hyperref}
\usepackage{oldgerm}
\usepackage{color}
\usepackage{bbm}

\newtheorem{theorem}{Theorem}[section]
\newtheorem{lemma}[theorem]{Lemma}
\newtheorem{corollary}[theorem]{Corollary}
\newtheorem{proposition}[theorem]{Proposition}
\newtheorem{definition}[theorem]{Definition}

\newtheorem{defi/prop}[theorem]{Definition/Proposition}

\def\squareforqed{\hbox{\rlap{$\sqcap$}$\sqcup$}}
\def\qed{\ifmmode\squareforqed\else{\unskip\nobreak\hfil
\penalty50\hskip1em\null\nobreak\hfil\squareforqed
\parfillskip=0pt\finalhyphendemerits=0\endgraf}\fi}

\newcommand{\N}{\mathbf{N}}

\newcommand{\R}{\mathbf{R}}
\newcommand{\C}{\mathbf{C}}
\renewcommand{\P}{\mathbf{P}}

\newcommand{\e}{\epsilon}
\renewcommand{\leq}{\leqslant}
\renewcommand{\geq}{\geqslant}

\newcommand{\st}{\  : \ }

\newcommand{\cH}{\mathcal{H}}

\newcommand{\cD}{\mathcal{D}}
\newcommand{\cP}{\mathcal{P}}
\newcommand{\cS}{\mathcal{S}}
\newcommand{\iy}{\infty}

\DeclareMathOperator{\vol}{vol}
\DeclareMathOperator{\vrad}{vrad}
\DeclareMathOperator{\conv}{conv}

\DeclareMathOperator{\tr}{Tr}
\DeclareMathOperator{\E}{\mathbf{E}}

\newcommand{\ketbra}[2]{| #1 \rangle \langle #2 |}
\newcommand{\bra}[1]{\langle #1 |}
\newcommand{\ket}[1]{| #1 \rangle}

\begin{document}
\title{Relaxations of separability in multipartite systems:\\Semidefinite programs, witnesses and volumes}

\author{C\'{e}cilia Lancien}
\affiliation{F\'{\i}sica Te\`{o}rica: Informaci\'{o} i Fenomens Qu\`{a}ntics, Universitat Aut\`{o}noma de Barcelona, 08193 Bellaterra, Barcelona, Spain}
\affiliation{Institut Camille Jordan, Universit\'{e} Claude Bernard Lyon 1, 43 boulevard du 11 novembre 1918, 69622 Villeurbanne Cedex, France}

\author{Otfried G\"uhne}
\affiliation{Naturwissenschaftlich-Technische Fakult\"at,
Universit\"at Siegen, Walter-Flex-Str. 3, 57068 Siegen, Germany}

\author{Ritabrata Sengupta}
\affiliation{Theoretical Statistics and Mathematics Unit, Indian Statistical Institute, 7 S.J.S. Sansanwal Marg, New Delhi 110 016, India}

\author{Marcus Huber}
\affiliation{F\'{\i}sica Te\`{o}rica: Informaci\'{o} i Fenomens Qu\`{a}ntics, Universitat Aut\`{o}noma de Barcelona, 08193 Bellaterra, Barcelona, Spain}
\affiliation{ICFO - Institut de Ciencies Fotoniques, 08860 Castelldefels, Barcelona, Spain}

\begin{abstract}
While entanglement is believed to be an important ingredient in
understanding quantum many-body physics, the complexity of its
characterization scales very unfavorably with the size of the
system. Finding super-sets of the set of separable states that
admit a simpler description has proven to be a fruitful
approach in the bipartite setting. In this paper we discuss a
systematic way of characterizing multiparticle entanglement
via various relaxations. We furthermore describe an
operational witness construction arising from such relaxations
that is capable of detecting every entangled state. Finally,
we also derive an analytic upper-bound on the volume of biseparable
states and show that the volume of the states with a positive partial
transpose for any split rapidly outgrows this volume.
This proves that simple semi-definite relaxations in the
multiparticle case cannot be an equally good approximation
for any scenario.
\end{abstract}

\maketitle

\section{Introduction}

Without a doubt entanglement can be considered one of the most important
concepts in quantum physics, clearly distinguishing quantum systems from
classical ones. It can be harnessed to enable novel ways of processing
quantum information in numerous ways, from communication to computation.
Many of these operational tasks require an operational detection or even
quantification of this indispensable resource. While in bipartite systems
of low dimensions this can be achieved in an efficient way, the complexity
of the characterization of entangled states makes a complete and computable
framework of entanglement detection impossible in high dimensions and thus
also for multipartite systems \cite{hororeview, gtreview}.

A possible way for deriving statements on the presence of entanglement
is to discard the actual complex structure of the border
between separable and entangled states and try to find good approximations
that admit a more amenable description. In the bipartite
case positive maps play a central role in such approximations: All separable
states remain positive semi-definite under application of a positive, yet not
completely positive map to one of its subsystems. The most well-known example of
such a map is the partial transposition: This map generally changes the
eigenvalues of a matrix, but separable states have a positive partial transpose (PPT). The approach of positive maps allows for a
characterization of super-sets of the set of separable states using techniques
from semi-definite programming, and it nevertheless captures the whole structure:
A state remains positive under {\it all} positive maps if and only if the state
is indeed separable \cite{hororeview, gtreview}.

In order to gauge the efficiency of certain maps for characterizing
entanglement, one of the most relevant issues is how the volume of
states that remain positive under the map in question compares to the
volume of separable states \cite{karol}. The sobering and non-surprising answer from
bipartite systems can be gained from convex geometry considerations and
shows that for all known maps the states that remain positive are most
likely entangled in high dimensions \cite{beigi}. For small dimensions,
however, a given positive map can detect a large fraction of entangled states.
This is, for instance, true for the partial transposition, which delivers
a necessary and sufficient criterion for $2 \times 2$ and $2 \times 3$ systems.

In multipartite systems the characterization of entanglement constitutes
an even greater challenge. Since partial separability of multipartite
states can no longer be defined as a purely bipartite concept, the application
of positive maps to subsystems alone can reveal little more than entanglement
across a fixed partition of the multipartite state. Nevertheless, recently
several works succeeded in defining suitable mixtures of positive maps, which
can be used to develop strong criteria for genuine multiparticle entanglement
\cite{PPTmixer, MHRBS}.

In this paper we first develop a framework that allows for the semi-definite
relaxation of partially separable states, opening the possibility for harnessing
well developed techniques based on positive maps to detect genuine multipartite
entanglement (an approach which has already been shown to yield useful
results with different relaxations in Ref.~\cite{doherty,eisert,brandao1,brandao2}).
We achieve this goal by first formally defining semi-definite
relaxations of partially separable states using positive maps.

Due to the formulation as semi-definite programs these constructions yield versatile
criteria for detecting multipartite entanglement in low dimensional systems. To unlock
these powerful techniques for more complex quantum states we proceed to discuss a recently
introduced program of lifting bipartite witnesses \cite{MHRBS}. We prove that it is always
possible to exploit witnesses that only reveal bipartite entanglement in order to construct
witnesses for genuine multipartite entanglement. This facilitates this notoriously hard problem, and
we showcase this technique with some exemplary multipartite entangled states.

In a second step, we ask which fraction of genuinely multipartite entangled states can be detected with such relaxation methods. We prove an upper-bound on the volume of the biseparable states, and a lower-bound on the volume of a set
of states that can never be detected with relaxation methods based on the partial
transposition. For large dimensions, both values deviate significantly. This shows
that while the relaxation approaches are strong for small systems, they
fail to deliver a good approximation in the general case.

\section{Characterizing relaxations of separability with semidefinite programs}
\label{section:SD relaxations}

The most straightforward relaxation of separability in multipartite systems
is again given by positive maps. Trying to justify this assertion is the
object of the current section.

Let us start with some basic definitions and notations (see e.g. Ref.~\cite{gtreview,ES} for a general review on these notions). On a multipartite
system, given a bipartition $\{b|\overline{b}\}$ of the
subsystems, we denote by $\mathcal{S}_b$ the set of states which are biseparable
across this cut, i.e.
\begin{equation}
\rho\in\mathcal{S}_b\ \text{if}\ \rho=\sum_i q_i|\phi_b^{(i)}\rangle\langle\phi_b^{(i)}|\otimes|\phi_{\overline{b}}^{(i)}\rangle\langle\phi_{\overline{b}}^{(i)}|,
\end{equation}
where $\{q_i\}_i$ is a convex combination, and for each $i$, $|\phi_b^{(i)}\rangle,|\phi_{\overline{b}}^{(i)}\rangle$ are pure states on the subsystems in $b,\overline{b}$ respectively.
We then define the set $\mathcal{S}_{(2)}$ of biseparable states as being the convex hull of $\{\mathcal{S}_b\}_b$, i.e.
\begin{equation}
\rho\in\mathcal{S}_{(2)}\ \text{if}\ \rho=\sum_{b}p_b\sigma_b,
\end{equation}
where $\{p_b\}_b$ is a convex combination, and for each $b$, $\sigma_b\in\mathcal{S}_b$. If a state is not biseparable it is called \emph{genuinely multipartite entangled} GME.

This definition admits a simple relaxation with a positive-semidefinite characterization. Given, for each bipartition $\{b|\overline{b}\}$, a positive map $\Lambda_b$, acting on the substystems in $b$, we define the set $\mathcal{R}_{\{\Lambda_b\}_b}$ of $\{\Lambda_b\}_b$-relaxation of $\mathcal{S}_{(2)}$ by
\begin{equation} \label{eq:relaxation} \rho\in\mathcal{R}_{\{\Lambda_b\}_b}\ \text{if}\ \rho=\sum_{b}p_b\sigma_{\Lambda_b}, \end{equation}
where $\{p_b\}_b$ is a convex combination and for each $b$, $\Lambda_b\otimes\mathbbm{1}_{\overline{b}}[\sigma_{\Lambda_b}]\geq 0$.

Such a relaxation carries the operational advantage that it
can be approached via semi-definite programming (SDP).
Besides, this definition can easily be extended to $k$-separable
states by applying different maps to the induced partitions.
As we are however mainly interested in characterizing the
strongest form of multipartite entanglement we focus here on the
distinction between biseparable states and genuinely
multipartite entangled ones.

For instance, these relaxations can be particularly useful when optimizing convex functions over the set of biseparable states. Indeed, for any function $f$, we trivially have that, for any set of positive maps $\{\Lambda_b\}_b$,
\begin{equation} \label{eq:optimize} \min_{\sigma\in\mathcal{S}_{(2)}}f(\sigma)\geq\min_{\sigma\in\mathcal{R}_{\{\Lambda_b\}_b}}f(\sigma).
\end{equation}
In particular, for any function $f$ satisfying $f(\lambda\rho_1+(1-\lambda)\rho_2)\leq \lambda f(\rho_1)+(1-\lambda)f(\rho_2)$, Eq.~\eqref{eq:optimize} provides a relaxation of the optimization of $f$ over $\mathcal{S}_{(2)}$ which can be cast as an SDP. One of the most straightforward applications of such a strategy is to testing whether a given density matrix $\rho$ is indeed biseparable, i.e. applying it e.g. to $f:\sigma\mapsto\left(\mathrm{Tr}\left[(\rho-\sigma)^2\right]\right)^{1/2}$. It yields in that case the equivalence
\begin{equation}
\forall\ \{\Lambda_b\}_b,\ \min_{\sigma\in\mathcal{R}_{\{\Lambda_b\}_b}}\mathrm{Tr}\big[\sigma(\sigma-2\rho)\big] \leq \mathrm{Tr}\left[\rho^2\right] \\ \Leftrightarrow\ \rho\in\mathcal{S}_{(2)}.
 \end{equation}

While this defines, in theory, a necessary and sufficient program for deciding whether a given state is biseparable, checking all possible sets of positive maps is of course not feasible. However even making one particular choice, such as the transposition map for instance, has already proven to yield very strong witnesses, and through the dual of the program one can additionally often extract analytical constructions for important classes of states \cite{PPTmixer}. To test the further prospects of such SDP approaches we have first programmed a Choi-map\cite{choilam} relaxation test in the following way:
\begin{align}
\max s\ \text{s.t.}\ \sigma_1\geq s\mathbbm{1},\, \sigma_2\geq s\mathbbm{1},\, \sigma_3=\rho-\sigma_1-\sigma_2,\, \sigma_1^{\mathcal{C}_1}\geq s\mathbbm{1},\, \sigma_2^{\mathcal{C}_2}\geq s\mathbbm{1},\, \sigma_3^{\mathcal{C}_3}\geq s\mathbbm{1}\,,
\end{align}
where $\mathcal{C}$ denotes the Choi map. While not being quantitative as the above program it can be easily implemented in MATLAB, using the packages YALMIP \cite{YALMIP} and the SEDUMI solver \cite{SEDUMI}. If an $s\geq0$ is found, the program also returns the decomposition elements $\sigma_i$ and can thus be used to find actual Choi-positive decompositions. As a first test of such a program we have chosen the family of states $\rho(\{\lambda_\alpha\})$ introduced in Ref.~\cite{MHRBS}. Setting all $\lambda_\alpha=\lambda$ it was shown in Ref.~\cite{MHRBS} that these states are GME for $0<\lambda<\frac{1}{3}$. Testing the program on this family of states returns a Choi-positive decomposition for $\frac{1}{3}\leq\lambda<1$, thus showing that on the set of Choi-positive mixtures the witness presented in Ref.~\cite{MHRBS} is weakly optimal. It furthermore proves that for $\frac{1}{3}\leq\lambda<1$ the state can be decomposed into states which are either PPT or Choi-positive.\\
We have also extended the program to demand simultaneous positivity under partial transposition and the Choi-map. That is formally, we looked for
\begin{align}
\max s\ \text{s.t.}\ \sigma_1\geq s\mathbbm{1},\,\sigma_2\geq s\mathbbm{1},\,\sigma_3=\rho-\sigma_1-\sigma_2,\, \sigma_1^{\mathcal{C}_1}\geq s\mathbbm{1},\,\sigma_2^{\mathcal{C}_2}\geq s\mathbbm{1},\,\sigma_3^{\mathcal{C}_3}\geq s\mathbbm{1},\nonumber\\ \sigma_1^{T_1}\geq s\mathbbm{1},\,\sigma_2^{T_2}\geq s\mathbbm{1},\,\sigma_3^{T_3}\geq s\mathbbm{1}\,.
\end{align}
With this program it is easy to check that indeed every state between $0<\lambda<1$ is genuinely multipartite entangled, which proves the versatility of this approach.

\section{Constructing multiparticle witnesses from bipartite witnesses}

While the semidefinite program presented before technically gives sufficient criteria for deciding biseparability, it becomes quickly intractable beyond a few qubits. There are, however, frequent situations in which one can use some additional knowledge to facilitate witness constructions.
Genuinely multipartite entangled states of course also have to be entangled across every bipartition of the system. And since the construction of bipartite entanglement witnesses can be a rather straightforward affair (e.g. through positive maps) one can ask whether there is a possibility to construct multipartite entanglement witnesses from a collection of bipartite operators.

\subsection{A systematic construction for lifting bipartite witnesses}

In Ref.~\cite{MHRBS} the authors introduced a general
witness construction method, which enables the construction
of multipartite entanglement witnesses from a set of
bipartite witnesses across every possible bipartition. Such
a problem can be formalized as follows.

Given, for each bipartition $\{b|\overline{b}\}$, a witness $W_b$, i.e. a self-adjoint operator with the property
that $\mathrm{Tr}(W_b\sigma_b)\geq0$ for all $\sigma_b\in\mathcal{S}_b$, we are looking for an
operator $W_{GME}$ with the following property
\begin{align} \forall\ b,\ W_{GME}\geq W_b, \label{general} \end{align}
where, for any self-adjoint operators $A,B$, we say that $A\geq B$ if $A-B$ is a positive semidefinite operator.

The construction of Ref.~\cite{MHRBS} constitutes one
analytic answer to this question in the following way.
Assume a common operator $Q$ for every bipartite
witness $W_b$, i.e.
\begin{equation}
\label{eq:W_b}
W_b=Q+T_b.
\end{equation}
Note that the only obstruction to the existence of such a non-trivial operator $Q$ is that the operators $\{W_b\}_b$ have the property that for each $b$ there exists $b'$ such that the non-zero eigenspace of $W_b$ is contained in the zero eigenspace of $W_{b'}$.
Otherwise, one can write down a formal solution of the above problem as
\begin{equation}
\label{eq:W_GME}
W_{GME}=Q+\sum_b\big[T_b\big]_+,
\end{equation}
where, for any self-adjoint operator $A$, we denote by $[A]_+$ the projection of $A$ onto the positive semidefinite cone.
It is now easy to see that condition
(\ref{general}) holds for every bipartite witness $W_b$.

Of course, the crucial issues here are first the one
of the optimal choice of $Q$ and second the one of the
generality of such a result. Using a special choice for $Q$,
the authors of Ref.~\cite{MHRBS} show that there exist
genuinely multipartite entangled states and a set of
bipartite witnesses for which the expectation value of
$W_{GME}$ is negative, proving that this construction indeed
succeeds in generating witnesses for multipartite
entanglement. Here we prove first of all that, for every
genuinely multipartite entangled state $\rho_{GME}$, there exist a set
of optimal bipartite witnesses for which this construction
yields a multipartite entanglement witness $W_{GME}$ such
that $\mathrm{Tr}(\rho_{GME}W_{GME})<0$.

\begin{theorem}
For every genuinely multipartite entangled state $\rho_{GME}$, there
exists a set of weakly optimal bipartite entanglement witnesses
$\{W_b\}_b$ such that
\begin{equation}
\mathrm{Tr}\left(\rho_{GME}\left(Q+\sum_b\big[T_b\big]_+\right)\right)<0,
 \end{equation}
where the operators $Q$ and $\{T_b\}_b$ are defined by equation \eqref{eq:W_b}.
\end{theorem}

\noindent\textit{Proof:}
Let $Q$ be a weakly optimal witness for multipartite entanglement, which means that $\forall\ \sigma\in\mathcal{S}_{(2)},\ \mathrm{Tr}(Q\sigma)\geq 0$, and $\exists\ b_0,\ \exists\ \rho_{b_0}\in\mathcal{S}_{b_0}:\ \mathrm{Tr}(Q\rho_{b_0})=0$. It is sufficient to show that such a witness
can always be constructed from bipartite weakly optimal witnesses, as then clearly
every multipartite entangled state can be detected by the construction.

{From} the assumption it follows that
\begin{equation}
\forall\ b,\ \min_{\rho_{b}\in\mathcal{S}_b}\mathrm{Tr}(Q\rho_{b})=:\alpha_{b}\geq 0.
\end{equation}
As this minimization is convex in the space of states it is clear that the minimal overlap with $Q$ can also be reached by an optimal pure state $|\psi_{b}\rangle\langle\psi_{b}|$. Now we can choose the following bipartite witness
\begin{equation}
 W_{b}=Q-\alpha_{b}|\psi_b\rangle\langle\psi_b|,
\end{equation}
and verify that, indeed, for each $b$, $\mathrm{Tr}(W_{b}\rho_{b})\geq0$ for all $\rho_b\in\mathcal{S}_b$ and $\mathrm{Tr}(W_{b}|\psi_{b}\rangle\langle\psi_{b}|)=0$, i.e.
that this operator is a weakly optimal witness for the bipartition $\{b|\overline{b}\}$.

Inserting this set of optimal bipartite
witnesses into the construction from above using $[-\alpha_{b}|\psi_b\rangle\langle\psi_b|]_+=0$ yields
\begin{equation}
 W_{GME}=Q,
 \end{equation}
proving that each weakly optimal multipartite witness can be gained
from the construction using weakly optimal bipartite witnesses. \hfill $\square$

\begin{corollary}
Every multipartite entanglement witness $W_{GME}$ can be constructed
using the framework summarized by equations \eqref{eq:W_b} and \eqref{eq:W_GME}. Furthermore, it is possible to impose that, for every bipartition $\{b|\overline{b}\}$,
$W_b=\Lambda^*_{b}\otimes\mathbbm{1}_{\overline{b}}\big[|\psi_b\rangle\langle\psi_b|\big]$ for some choice of positive map $\Lambda_{b}$, acting on the subsystems in $b$, and some choice of pure state $|\psi_b\rangle$.
\end{corollary}

\noindent\textit{Proof:}
First it is important to notice that, for every state $\rho$ which is
entangled across a specific bipartition $\{b|\overline{b}\}$,
there exists a positive map $\Lambda_b$, acting on the subsystems in $b$, such that
$\Lambda_b\otimes\mathbbm{1}_{\overline{b}}[\rho]\ngeq 0$,
i.e. it is detected to be entangled by this positive map
(see \cite{hororeview, depillis, jamiolkowski}). This implies in turn that there exists a
pure state $|\psi_b\rangle$ such that $\langle\psi_b|\Lambda_b\otimes\mathbbm{1}_{\overline{b}}[\rho]|\psi_b\rangle< 0$, i.e.
\begin{equation}
 \mathrm{Tr}\left(\rho\Lambda^*_{b}\otimes\mathbbm{1}_{\overline{b}}\big[|\psi_b\rangle\langle\psi_b|\big]\right)<0.
\end{equation}
Through continuity this implies that, for every extremal biseparable state $|\phi\rangle$, there exists a weakly optimal witness of the form $\widetilde{W}_b=\Lambda^*_{b}\otimes\mathbbm{1}_{\overline{b}}\big[|\psi_b\rangle\langle\psi_b|\big]$ such that $\mathrm{Tr}\big(\widetilde{W}_b|\phi\rangle\langle\phi|\big)=0$. Now, invoking the Choi-Jamiolkowski map, we can conclude that every possible hyperplane intersecting the biseparable set corresponds to a witness derived from a positive map. This implies that amongst all the $\{\widetilde{W}_b\}_b$, there is at least one them $\widetilde{W}_{b_0}$ which satisfies the following: for each bipartition $\{b|\overline{b}\}$, $W_b=\widetilde{W}_{b_0}-\beta_b\mathbbm{1}$, for some $\beta_b$, is an optimal bipartite witness for $\rho$ (and $\beta_{b_0}=0$). This concludes the proof: the $\{W_b\}_b$ are all obtained from shifting $\widetilde{W}_{b_0}=\Lambda^*_{b_0}\otimes\mathbbm{1}_{\overline{b_0}}\big[|\psi_{b_0}\rangle\langle\psi_{b_0}|\big]$. \hfill $\square$

\subsection{Illustration on one example}

To showcase the strength of the above constructions let us illustrate it with a peculiar example. First of all let us make a specific choice for $Q$ that has already proven to work well in Ref.~\cite{MHRBS}: Given a set of witnesses for every bipartition $\{W_b\}_b$, we will construct $Q$ by finding element wise the matrix of largest common negative matrix entries
\begin{align}
N=\sum_{\eta,\eta'\in\{0,1,...,d-1\}}|\eta\rangle\langle\eta'|\min\left[0,\max_{b}\big[\Re e\bra{\eta}W_b\ket{\eta'}\big]\right]\,,
\end{align}
and the matrix of smallest common positive matrix entries
\begin{align}
P=\sum_{\eta,\eta'\in\{0,1,...,d-1\}}|\eta\rangle\langle\eta'|\max\left[0,\min_{b}\big[\Re e\bra{\eta}W_b\ket{\eta'}\big]\right]\,.
\end{align}
With these two matrices we can define
\begin{align}
Q:=N+P\,.
\end{align}
Now for the purpose of elucidating how this construction works in practice let us follow it through step by step in an exemplary case. To define our target state a few shorthand notations will be useful. Fix $a,b,c>0$ and define the following vectors in a three-qutrit system
\begin{align}
& |\psi_1\rangle=\sqrt{a} |001\rangle+\sqrt{a^{-1}}|110\rangle,
& |\psi_2\rangle=\sqrt{a} |010\rangle+\sqrt{a^{-1}}|101\rangle,
\nonumber \\
& |\psi_3\rangle=\sqrt{a} |100\rangle+\sqrt{a^{-1}}|011\rangle,
& |\psi_4\rangle=\sqrt{b} |112\rangle+\sqrt{b^{-1}}|221\rangle,
\nonumber\\
& |\psi_5\rangle=\sqrt{b} |121\rangle+\sqrt{b^{-1}}|212\rangle,
& |\psi_6\rangle=\sqrt{b} |211\rangle+\sqrt{b^{-1}}|122\rangle,
\nonumber\\
& |\psi_7\rangle=\sqrt{c} |220\rangle+\sqrt{c^{-1}}|002\rangle,
& |\psi_8\rangle=\sqrt{c} |202\rangle+\sqrt{c^{-1}}|020\rangle,
\nonumber\\
& |\psi_9\rangle=\sqrt{c} |022\rangle+\sqrt{c^{-1}}|200\rangle,
& |\psi_{10}\rangle=|000\rangle+|111\rangle+|222\rangle.
\end{align}
We can then construct a mixed state $\rho$ as follows
\begin{align}
& \tilde{\rho}=\sum_{i=1}^{10}|\psi_i\rangle\langle\psi_i|+p(|001\rangle\langle 001|+\mathbbm{1})
\nonumber\\
& \rho:=\frac{\tilde{\rho}}{\text{Tr}\,\tilde{\rho}}\,.
\end{align}
By construction this state is PPT across all three cuts. If we choose $a=10^{-6}$, $b=300$ and $c=12\times10^{-3}$, then for $p>0.0003$ it is also positive under application of the Choi map (or its dual) across all three cuts. As it is both PPT and Choi-positive it is fair to say that if it is entangled it is only very weakly so. The system size is still small enough to apply the map-mixer introduced in the previous section, and to reveal that it is nonetheless indeed genuinely multipartite entangled for various values of $p>0.003$.

Hence, this is a good opportunity to demonstrate the power of witness liftings. We can use the indecomposable map introduced in Ref.~\cite{OSA}, choosing $c_1=1$, $c_2=10^{-3}$ and $c_3=10^3$, revealing entanglement across all three cuts. In fact, even without ever calculating any eigenvalues, we can just apply the map on the three subsystems of $|\psi_{10}\rangle$ only. That is
\begin{align}
& W_1=\Lambda^{c_1,c_2,c_3}_1\otimes\mathbbm{1}_{23}[|\psi_{10}\rangle\langle\psi_{10}|], \nonumber \\
& W_2=\Lambda^{c_1,c_2,c_3}_2\otimes\mathbbm{1}_{13}[|\psi_{10}\rangle\langle\psi_{10}|], \nonumber \\
& W_3=\Lambda^{c_1,c_2,c_3}_3\otimes\mathbbm{1}_{12}[|\psi_{10}\rangle\langle\psi_{10}|].
\end{align}
Plugging the three resulting witnesses in our construction for $Q$ we get a GME-witness that is able to reveal genuine multipartite entanglement in the state for a range of $0\leq p\leq 0.00069$. This example illustrates that even bound entangled states which are positive with respect to paradigmatic maps can exhibit multipartite entanglement. As the state itself is full rank and not symmetric, there is no other known method that could have revealed it to be GME, clearly demonstrating the power of positive map mixers and witness liftings.

\subsection{Finding a multiparticle witness with semidefinite
programming}
In Ref.~\cite{MHRBS} the multipartite witness $W_{GME}$ was constructed
by starting with bipartite witnesses $W_b$ for each bipartition
$\{b|\overline{b}\}$, and then one possible choice for the operator
$Q$ was explicitly  constructed. While the presented choice
works well for many examples, as illustrated in the previous section,
it is not clear whether it is optimal in the general case. For a given state $\rho$, however, the optimal
multipartite witness $W_{GME}$ can directly be computed as a semi-definite program that is easier to run than the map-mixers.

For that, consider the following constrained optimization problem
\begin{align}
\mbox{ \text{minimize:} } \quad&  \mathrm{Tr}\big(\rho W_{GME}\big)
\\
\mbox{ \text{subject to:} } \quad& \forall\ b,\ W_{GME} \geq W_b.
\end{align}
This is a semidefinite program, which can be easily and efficiently solved
using standard numerical techniques.

There is, in addition, a variation of the presented method for obtaining
a multiparticle witness from a set of bipartite witnesses. The condition
imposed by Eq.~(\ref{general}) guarantees that, for each $b$,
$\mathrm{Tr}(\sigma W_{GME}) \geq \mathrm{Tr}(\sigma W_{b})$ for any
state $\sigma$, so that $W_{GME}$ is indeed an entanglement witness.
However, for being an entanglement witness it suffices that, for
each $b$, $\mathrm{Tr}(\sigma_b W_{GME}) \geq \mathrm{Tr}(\sigma_b W_{b})$
for any state $\sigma_b$ which is separable for the bipartition
$\{b|\overline{b}\}$. And this is already guaranteed if, for all
$b$, there exists a positive map $\Lambda_b$, acting on subsystems
in $b$, such that
\begin{equation} \Lambda_b \otimes \mathbbm{1}_{\overline{b}} \big[W_{GME}\big] \geq \Lambda_b \otimes \mathbbm{1}_{\overline{b}} \big[W_{b}\big].
\end{equation}
So, for computing a multipartite witness
for a given state one can also consider the semidefinite program
\begin{align}
\mbox{ \text{minimize:} } \quad&  \mathrm{Tr}\big(\rho W_{GME}\big)
\\
\mbox{ \text{subject to:} } \quad& \forall\ b,\ \Lambda_b \otimes \mathbbm{1}_{\overline{b}} \big[W_{GME}\big] \geq \Lambda_b \otimes \mathbbm{1}_{\overline{b}} [W_b].
\end{align}
These two semidefinite programs can be used to construct the best witness for a given
state systematically. They might be useful, if the analytical method from Ref.~\cite{MHRBS}
does not work.

It should be noted, however, that the presented formulations are not necessarily
the best way to construct a multiparticle witness from  bipartite witnesses, if
one wishes to use semidefinite programming. The reason is the following: For each
bipartite witness one can construct  via the Choi isomorphism a positive, but not
completely positive map. This map detects more states than the original witness.
Given these maps one can then evaluate the corresponding map relaxations, as described
in Section 2. This criterion will be stronger than the semidefinite programs presented
above. For that reason, we do not discuss detailed examples here.

\section{Relaxations of separability beyond positive maps}
\label{section:CCNR}

So far, we considered only the approximation of the biseparable sets by
super-sets which are associated to a positive map. One can, however, also
use other separability criteria, such as the computable cross norm, aka
realignment, (CCNR) criterion \cite{ccnr1, ccnr2} or symmetric extensions
\cite{doherty}. In the following we explain how the CCNR criterion can be used.

\subsection{Description of the method}

Let us start by explaining the CCNR criterion. Any bipartite quantum
state can be expressed via the Schmidt decomposition in operator space
as
\begin{equation}
\rho_{AB} = \sum_k \lambda_k G^A_k \otimes G^B_k,
\end{equation}
where the $\lambda_k$ are positive coefficients, and the $G^A_k$ and $G^B_k$
are orthogonal observables, i.e. they fulfill $\mathrm{Tr}(G^A_i G^A_j)= \mathrm{Tr}(G^B_i G^B_j)= \delta_{ij}$. With this representation one can easily prove that the
following holds:
\begin{equation}
\rho_{AB} \mbox{ separable } \Rightarrow  \sum_k \lambda_k \leq 1.
\end{equation}
And this necessary condition for separability is known as the CCNR criterion.
The criterion has the advantage that it detects entanglement in many states
where the PPT criterion fails. On the other hand, not all two-qubit entangled
states can be detected by this test.

{From} this structure, one can easily write down entanglement witnesses.
Namely, any operator of the form
\begin{equation}
W = \mathbbm{1} - \sum_k G^A_k \otimes G^B_k
\end{equation}
is an entanglement witness, as it is positive on all states with $\sum_k \lambda_k \leq 1$.
This structure of the witnesses can be used for constructing witnesses for
genuine multiparticle entanglement as follows. Consider a witness which has
for any bipartition $\{b|\bar b\}$ the structure
\begin{align}
W_{GME} =  P_b + \mathbbm{1} - \sum_k G^{b}_k \otimes G^{\bar b}_k,
\label{ccnrcond}
\end{align}
where $G^{b}_k$ and $G^{\bar b}_k$ are orthogonal observables on
the bipartition $\{b|\bar b\}$, and $P_b\geq 0$ is positive semidefinite.
Clearly, if a state obeys the CCNR criterion for some bipartition,
the mean value of the witness $W_{GME}$ will not be negative.
Consequently, the witness is also non-negative on all biseparable
states.

\subsection{Example: The three-qubit GHZ state}
The witnesses from the CCNR criterion are more difficult to handle
than the witnesses from positive maps. The reason is that no approach
via semidefinite programming is possible. Moreover, the condition from
Eq.~(\ref{ccnrcond}) is more difficult to check than the condition in
Eq.~(\ref{general}). Nevertheless, we will present an example where
known optimal entanglement witnesses have this structure.

Consider first the three-qubit Greenberger-Horne-Zeilinger (GHZ)
state $\ket{GHZ}= (\ket{000}+\ket{111})/\sqrt{2}.$
The typical witness for this state  is $W= \mathbbm{1}/2 - \ketbra{GHZ}{GHZ}$.
Now, the GHZ state can be expressed in terms of its stabilizers as
\begin{equation} \ketbra{GHZ}{GHZ} = \frac{1}{8}(111 + ZZ1 + Z1Z + 1ZZ + XXX - XYY - YXY - YYX),
\end{equation}
where $1,X,Y,Z$ represent the Pauli matrices $\mathbbm{1}, \sigma_x, \sigma_y, \sigma_z$,
and tensor product signs have been omitted. After a change of the normalization
this can be used to write the witness as
\begin{align}
W &= {\mathbbm{1}} - 2 \ketbra{GHZ}{GHZ}
\\
&=  \mathbbm{1} - \left[\frac{Z}{\sqrt{2}} \frac{Z1+1Z}{\sqrt{8}} +
\frac{X}{\sqrt{2}} \frac{XX-YY}{\sqrt{8}} +
\frac{-Y}{\sqrt{2}} \frac{YX+XY}{\sqrt{8}} +
\frac{1}{\sqrt{2}} \frac{11+ZZ}{\sqrt{8}}\right].
\end{align}
{From} this representation, it is clear that $W$ is a witness as in
Eq.~(\ref{ccnrcond}) for the bipartition $\{\{1\}|\{2,3\}\}$, with $P_1=0.$
Due to the symmetry, this works for all bipartitions.

\section{Estimating the performance of PPT relaxations in high dimensions}
\label{section:volumes}
In this final section we want to discuss the overall performance of such relaxations in multipartite systems, using the paradigmatic partial transpose map. In order to estimate the performance of using PPT relaxations to detect randomly chosen multipartite entangled states we derive lower-bounds on the fraction of multipartite entangled states, among states which are positive under partial transposition across every cut. The latter condition is strictly stronger than the relaxation employed in Eq.~\eqref{eq:relaxation}, thus providing an upper-bound on the fraction of states in $\mathcal{R}_{\{T_b\}_b}$ (where $T_b$ stands for the transposition on the subsystems in $b$) that are also in $\mathcal{S}_{(2)}$.

Our main result can be summarized as follows: \emph{For a fixed number of parties, the ratio between the volume of biseparable states and the volume of fully PPT states (as measured by either the volume radius or the mean width) scales as $1/\sqrt{d}$, where $d$ is the local dimension.}

In order to precisely formulate this result, we need to introduce first some of the basic notions and definitions that will be employed in the derivations.

\subsection{Notation and preliminary technical remarks}

We denote by $\cH(\C^n)$ the set of Hermitian operators on
$\C^n$, on which we define $\|\cdot\|_{tr}$ as the trace
class norm, $\|\cdot\|_{HS}$ as the Hilbert--Schmidt norm
and $\|\cdot\|_{op}$ as the operator norm. We also denote by
$\|\cdot\|$ the Euclidean norm on $\C^n$. In what follows, we will be mostly interested in the asymptotic regime, when the dimension $n$ tends to infinity. In this setting, the letters $C,c,c_0$ etc. denote numerical constants, independent from any other parameters, in particular from $n$. The value of these constants may change from occurrence to occurrence. When $A$ and $B$ are quantities depending on $n$, the notation $A \sim B$ means that the ratio $A/B$ tends to $1$ when $n$ tends to infinity.

When working with a random variable $X$, we will use the notation $\P(\mathcal{A}(X))$ to denote the probability of the event $\mathcal{A}(X)$, and the notation $\E\left(f(X)\right)$ to denote the expectation of the function $f(X)$.

Extra notation, concepts and results from classical convex
geometry, which are required throughout our proofs, are
gathered in Appendix \ref{ap:convex-geometry}.

It may, however, be worth mentioning that whenever we use
tools from convex geometry in the space $\cH(\C^n)$ (which
has real dimension $n^2$) it is tacitly understood that we
use the Euclidean structure induced by the Hilbert--Schmidt
inner product $\langle A,B \rangle = \tr(AB)$. For instance,
Definition \ref{def:vrad} of the volume radius of a convex
body $K\subset\cH(\C^n)$ becomes, denoting by $B_{HS}$ the
Hilbert--Schmidt unit ball of $\cH(\C^n)$,
\begin{align} \vrad (K) = \left( \frac{\vol K}{\vol B_{HS}}
\right)^{1/n^2}. \end{align}
While Definition \ref{def:w} of its mean width is, denoting
by $S_{HS}$ the Hilbert--Schmidt unit sphere of $\cH(\C^n)$
equipped with the uniform probability measure $\sigma$,
\begin{align} w (K) = \int_{S_{HS}} \underset{M\in K}{\max}\tr(X M) \, \mathrm{d} \sigma (X).  \end{align}
As also mentioned in Appendix \ref{ap:convex-geometry}, the
latter quantity can be re-expressed via Gaussian variables,
which yields here
\begin{align} w(K) = \frac{1}{\gamma_n} \E \left(\underset{M\in K}{\max}\tr(G M)\right), \end{align}
where $G$ is a matrix from the Gaussian Unitary Ensemble (GUE) on $\C^n$ and $\gamma_n=\E\|G\|_{HS}\sim_{n\rightarrow+\iy}n$ (see e.g. Ref.~\cite{AGZ}, Chapter 2, for a proof).

To be fully rigorous, let us make one last comment. All the
convex bodies of $\cH(\C^n)$ that we shall consider will
actually be included in the set $\cD(\C^n)$ of density
operators on $\C^n$ (i.e. the set of positive and trace $1$
operators on $\C^n$). So we will in fact be working in an
ambient space of real dimension $n^2-1$ (namely the
hyperplane of $\cH(\C^n)$ composed of trace $1$ elements).
This subtlety will not be an issue though, since we will be
mostly interested in the asymptotic regime
$n\rightarrow+\iy$. In this setting, the operator that will
play for us the role of the origin will naturally be the
center of mass of $\cD(\C^n)$, i.e. the maximally mixed
state $\mathbbm{1}/n$.


\begin{theorem} \label{th:vrad-w-states}
On $\C^n$, the volume radius and the mean width of the set $\cD$ of all quantum states satisfy the asymptotic estimates,
\begin{equation} \label{eq:vrad-states} \vrad(\cD) \underset{n\rightarrow+\iy}{\sim}\frac{e^{-1/4}}{\sqrt{n}}, \end{equation}
\begin{equation} \label{eq:w-states} w(\cD) \underset{n\rightarrow+\iy}{\sim}\frac{2}{\sqrt{n}}. \end{equation}
\end{theorem}

\noindent\textit{Proof:}
Eq.~\eqref{eq:vrad-states} was established in Ref.~\cite{SZ}.

Eq.~\eqref{eq:w-states} is a direct consequence of Wigner's semicircle law (see e.g. Ref.~\cite{AGZ}, Chapter 2, for a proof). Indeed, we have by definition
\begin{align} w(\cD) = \frac{1}{\gamma_n}\E \left(\underset{\rho\in\cD}{\max}\tr\left[G\left(\rho-\frac{\mathbbm{1}}{n}\right)\right]\right)
= \frac{1}{\gamma_n}\E \left(\underset{\rho\in\cD}{\max}\tr[G\rho]\right)
= \frac{1}{\gamma_n} \E \left(\lambda_{max}(G)\right), \end{align}
where G is a GUE matrix on $\C^n$ and we denoted by $\lambda_{max}(G)$ its largest eigenvalue. The claimed result then follows from $\gamma_n\sim_{n\rightarrow+\infty}n$ and $\E \left(\lambda_{max}(G)\right)\sim_{n\rightarrow+\infty}2\sqrt{n}$. \hfill $\square$

\subsection{Volume estimates}

In the sequel, we shall consider the multipartite system $(\C^d)^{\otimes k}$, and slightly adapt and generalize the notations introduced in Section \ref{section:SD relaxations}. We shall denote by $\cS$ and $\cP$ the sets of states on $(\C^d)^{\otimes k}$ which are, respectively, separable and PPT across any bi-partition, and by $\cS_{(2)}$ and $\cP_{(2)}$ the sets of states on $(\C^d)^{\otimes k}$ which are, respectively, bi-separable and bi-PPT. These sets may be more precisely defined in the following way. There are $N_k=2^{k-1}-1$ different bi-partitions of the $k$ subsystems $\C^d$. Denoting by $\left\{\cS^1,\ldots,\cS^{N_k}\right\}$ and by $\left\{\cP^1,\ldots,\cP^{N_k}\right\}$ the sets of states which are, respectively, bi-separable and bi-PPT across one of these, we have
\begin{align*} \cS=\bigcap_{i=1}^{N_k}\cS^i\ \ \text{and} & \ \ \cP=\bigcap_{i=1}^{N_k}\cP^i,\\
\cS_{(2)}=\conv\left(\bigcup_{i=1}^{N_k}\cS^i\right)\ \ \text{and} & \ \ \cP_{(2)}=\conv\left(\bigcup_{i=1}^{N_k}\cP^i\right). \end{align*}

\begin{theorem} \label{th:fully-ppt} There exist positive constants $c_d\rightarrow_{d\rightarrow+\infty}1$ such that, on $(\C^d)^{\otimes k}$, the volume radius and the mean width of the set of states which are PPT across any bi-partition satisfy
\begin{equation} \label{eq:vrad-w-ppt} w(\cP)\geq\vrad(\cP)\geq c_d\, \frac{c^{2^k}}{d^{k/2}}, \end{equation}
where one may choose $c=e^{-1/4}/4$.
\end{theorem}

\noindent\textit{Proof:}
The first inequality in Eq.~\eqref{eq:vrad-w-ppt} is just by the Urysohn inequality (see Theorem \ref{th:urysohn}).

To show the second inequality in Eq.~\eqref{eq:vrad-w-ppt}, we will use repeatedly the Milman--Pajor inequality (see Theorem \ref{th:Milman-Pajor}) and more specifically its Corollary \ref{corollary:Milman-Pajor}. We will in fact show more precisely that there exist $c_d\rightarrow_{d\rightarrow+\infty}1$ such that
\begin{equation}\label{eq:vrad-ppt} \vrad(\cP)\geq c_d\,\frac{c^{N_k}\,e^{-1/4}}{d^{k/2}}. \end{equation}
The first thing to note is that, denoting by $\Gamma_1,\ldots,\Gamma_{N_k}$ the partial transpositions across the $N_k$ different bi-partitions of the $k$ subsystems $\C^d$, we have
\begin{align} \cP = \cD\cap\cD^{\Gamma_1}\cap\cdots\cap\cD^{\Gamma_{N_k}}. \end{align}
Now, by Corollary \ref{corollary:Milman-Pajor} applied to the convex body $\cD\subset\cH((\C^d)^{\otimes k})$ (which indeed has the origin $\mathbbm{1}/d^k$ as center of mass) and to the isometry $\Gamma_1$, we get
\begin{align} \vrad\left(\cD\cap\cD^{\Gamma_1}\right)\geq\frac{1}{2}\frac{\vrad(\cD)^2}{w(\cD)} \underset{d\rightarrow+\iy}{\sim}c\times\frac{e^{-1/4}}{d^{k/2}}, \end{align}
the last equivalence being by Theorem \ref{th:vrad-w-states}. We may then conclude recursively that Eq.~\eqref{eq:vrad-ppt} actually holds. \hfill $\square$

\medskip

\begin{theorem} \label{th:bi-sep} On $(\C^d)^{\otimes k}$, the volume radius and the mean width of the set of bi-separable states satisfy
\begin{equation} \label{eq:vrad-w-sep2} \vrad(\cS_{(2)})\leq w(\cS_{(2)})\leq \frac{C+C_{d,k}}{d^{(k+1)/2}}, \end{equation}
where one may choose $C=\min\left\{6\sqrt{\ln(1+2/\delta)}/(1-2\delta^2)^2 \st 1/10<\delta<1/4 \right\}$ and $C_{d,k}=\sqrt{8\ln(2)/d^{k-1}}$, so that $C\leq 11$ and $C_{d,k}\rightarrow_{d\rightarrow+\infty}0$.
\end{theorem}

\noindent\textit{Proof:}
The first inequality in Eq.~\eqref{eq:vrad-w-sep2} is just by the Urysohn inequality (see Theorem \ref{th:urysohn}).

To show the second inequality in Eq.~\eqref{eq:vrad-w-sep2}, we start from the following observation: Let $\widetilde{\cS}$ be one of the $k$ set of states on $(\C^d)^{\otimes k}$ which are separable across a given cut $\C^d:(\C^d)^{\otimes k-1}$. Then, for each $1\leq i\leq N_k$,
\begin{equation}\label{eq:w(S^i)} w(\cS^i)\leq w(\widetilde{\cS})\leq \frac{C/2}{d^{(k+1)/2}}, \end{equation}
where $C=\min\big\{6\sqrt{\ln(1+2/\delta)}/(1-2\delta^2)^2 \st 1/10<\delta<1/4 \big\}$. It relies on the already known fact that there exists a universal constant $\widetilde{C}$ such that, for any $m,n\in\N$ with $m\leq n$, the mean width of the set $\cS$ of separable states on $\C^m\otimes\C^n$ is upper-bounded by $\widetilde{C}/m\sqrt{n}$. This result was basically proved in Ref.~\cite{AS}, Theorem 1, but since specifically stated there in the balanced case $m=n$ only, for $\vrad(\cS)$ rather than $w(\cS)$ and without specifying that one may choose $\widetilde{C}=C/2$, we briefly recall the argument here.

Let $1/10<\delta<1/4$ and consider $\mathcal{M}_{\delta}$, $\mathcal{N}_{\delta}$ $\delta$-nets for $\|\cdot\|$ within the Euclidean unit spheres of $\C^m$ and $\C^n$ respectively. Imposing that $\mathcal{M}_{\delta}$, $\mathcal{N}_{\delta}$ have minimal cardinality, we know by volumetric arguments (see e.g. Ref.~\cite{Pisier}, Lemma 4.16) that $\left|\mathcal{M}_{\delta}\right|\leq\left(1+2/\delta\right)^{2m}$ and $\left|\mathcal{N}_{\delta}\right|\leq\left(1+2/\delta\right)^{2n}$. Then, it may be checked that
\begin{align} \conv\left(\cS\cup-\cS\right)\subset\frac{1}{(1-2\delta^2)^2}\conv\big\{\pm\ket{x}\bra{x}\otimes\ket{y}\bra{y} \st \ket{x}\in\mathcal{M}_{\delta},\ \ket{y}\in\mathcal{N}_{\delta}\big\}. \end{align}
So by Lemma \ref{lemma:mean-width-polytope}, we get
\begin{align} w(\cS) \leq\, & w\left(\conv\left(\cS\cup-\cS\right)\right)\\
\leq\, & \frac{1}{(1-2\delta^2)^2}\sqrt{\frac{2\ln\left(2(1+2/\delta)^{2m}(1+2/\delta)^{2n}\right)}{(mn)^2}}\\
=\, & \frac{1}{(1-2\delta^2)^2}\frac{\sqrt{4(m+n)\ln\left(1+2/\delta\right)+\ln(4)}}{mn}\\
\leq\, & \frac{3\sqrt{\ln\left(1+2/\delta\right)}}{(1-2\delta^2)^2m\sqrt{n}}, \end{align}
which is precisely the content of Eq.~\eqref{eq:w(S^i)}.

Now, we also have that, for each $1\leq i\leq N_k$, $\cS^i\subset B_{tr}\subset B_{HS}$. Hence, by Lemma \ref{lemma:mean-width-union-bound}, we get
\begin{align} w(\cS_{(2)}) \leq 2 \left( \underset{1\leq i\leq N_k}{\max} w(\cS^i) + \sqrt{\frac{2\ln(N_k)}{(d^k)^2}} \right) \leq \frac{C}{d^{(k+1)/2}} + \frac{\sqrt{8\ln(2)k}}{d^k}=\frac{C+C_{d,k}}{d^{(k+1)/2}}, \end{align}
where $C_{d,k}=\sqrt{8\ln(2)/d^{k-1}}$. \hfill $\square$

\medskip

The conclusion of Theorems \ref{th:fully-ppt} and \ref{th:bi-sep} may be phrased as follows. On a multipartite system which is composed of a small number of big subsystems ($k$ fixed and $d\rightarrow+\infty$), imposing that a state is PPT across any bi-partition (i.e. the strongest notion of PPT) is still, on average, a much less restrictive constraint than imposing that it is bi-separable (i.e. the weakest notion of separability). Indeed, the ``sizes'' of these two sets of states (measured by either their volume radii or their mean widths) scale completely differently: the ``size'' of the former is at least of order $1/d^{k/2}$ while the ``size'' of the latter is at most of order $1/d^{(k+1)/2}$, hence differing by a factor of order at least $\sqrt{d}$.

\subsection{A class of fully PPT and GME states}

In Ref.~\cite{MHRBS} an explicit class of GME states that were PPT across all cuts was presented. In small dimensions it is a hard task to find such examples, but the results from the previous section suggest that at least in high dimensions this should be a generic feature of PPT states. To emphasize this fact we present an explicit construction of states PPT across all cuts and GME with high probability.

Consider the following random state model on $(\C^d)^{\otimes k}$: fix some parameter $0<\alpha<1/4$ (independent of $d$), pick $G$ a traceless GUE matrix on $(\C^d)^{\otimes k}$, and define the ``maximally mixed + gaussian noise'' state on $(\C^d)^{\otimes k}$
\begin{equation}\label{eq:rho_G} \rho_G=\frac{1}{d^k}\left(\mathbbm{1}+\frac{\alpha}{d^{k/2}}G\right). \end{equation}
Then, typically (i.e. with probability going to $1$ as $d$ grows) $\rho_G$ is fully PPT and nevertheless GME.

More quantitatively, we will show that the following result holds.

\begin{theorem} \label{th:fullyPPT-GME}
Let $G$ be a traceless GUE matrix on $(\C^d)^{\otimes k}$. Then, the state $\rho_G$ on $(\C^d)^{\otimes k}$, as defined by Eq.~\eqref{eq:rho_G}, is fully PPT and not bi-separable with probability greater than $1-\exp(-cd^{k-1})$, for some universal constant $c>0$.
\end{theorem}

Theorem \ref{th:fullyPPT-GME} is a straightforward consequence of Propositions \ref{prop:fullyPPT} and \ref{prop:GME} below. Before stating and proving them, let us elude once and for all a slight issue: a GUE matrix on $\C^n$ is the standard Gaussian vector in $\cH(\C^n)$, while a traceless GUE matrix on $\C^n$ is the standard Gaussian vector in the hyperplane of $\cH(\C^n)$ composed of trace $0$ elements. So in the asymptotic regime $n\rightarrow+\infty$, all the known results on $n\times n$ GUE matrices that we shall use also hold for traceless $n\times n$ GUE matrices (because the ambient spaces of these two gaussian vectors have equivalent dimensions in this limit).

\begin{proposition} \label{prop:fullyPPT}
Let $G$ be a traceless GUE matrix on $(\C^d)^{\otimes k}$. Then, the state $\rho_G$ on $(\C^d)^{\otimes k}$, as defined by Eq.~\eqref{eq:rho_G}, satisfies
\begin{align} \P\left( \rho_G\notin\cP\right) \leq e^{-cd^k}, \end{align}
where $c>0$ is a universal constant.
\end{proposition}

\noindent\textit{Proof:} In Ref.~\cite{Aubrun}, a deviation inequality is proved for the smallest eigenvalue of a GUE matrix, namely: Let $G$ be a GUE matrix on $\C^n$ and denote by $\lambda_{min}(G)$ its smallest eigenvalue. Then, for any $\e>0$,
\begin{align} \P\left(\lambda_{min}(G)<-(2+\e)\sqrt{n}\right) \leq e^{-c\e^{3/2}n}, \end{align}
where $c>0$ is a universal constant.

Now, observe that $G$ as well as all its partial transpositions $G^{\Gamma_i}$, $1\leq i\leq N_k$, are GUE matrices on $(\C^d)^{\otimes k}$. Hence, Proposition \ref{prop:fullyPPT} follows directly, by choosing for instance $\e=1$. Indeed, by assumption on $\alpha$, we have $3\alpha<3/4<1$, so the probability that $\rho_G$ or any $\rho_G^{\Gamma_i}$, $1\leq i\leq N_k$, is not positive is less than $e^{-cd^k}$. \hfill $\square$

\medskip

\begin{proposition} \label{prop:GME}
Let $G$ be a traceless GUE matrix on $(\C^d)^{\otimes k}$. Then, the state $\rho_G$ on $(\C^d)^{\otimes k}$, as defined by Eq.~\eqref{eq:rho_G}, satisfies
\begin{align} \P\left( \rho_G\in\cS_{(2)}\right) \leq e^{-cd^{k-1}}, \end{align}
where $c>0$ is a universal constant.
\end{proposition}

\noindent\textit{Proof:} Our strategy to show Proposition \ref{prop:GME} is to exhibit a Hermitian $M$ on $(\C^d)^{\otimes k}$ which is with probability greater than $1-\exp(-cd^{k-1})$ a GME witness for the state $\rho_G$ (i.e. such that $\tr(\rho_G M)<0$ while $\tr(\rho M)>0$ for any bi-separable state $\rho$).

Note first of all that, on the one hand,
\begin{equation} \label{eq:E1} \E\tr(\rho_GG) = \frac{\alpha}{d^{3k/2}}\E\tr(G^2) \underset{d\rightarrow +\infty}{\sim} \frac{\alpha}{d^{3k/2}}d^{2k} = \alpha d^{k/2}, \end{equation}
while on the other hand, by Theorem \ref{th:bi-sep},
\begin{equation} \label{eq:E2} \E\underset{\rho\in\cS_{(2)}}{\sup}\tr(\rho G) \leq \frac{C+C_{d,k}}{d^{(k+1)/2}} \E[\tr(G^2)]^{1/2} \underset{d\rightarrow +\infty}{\sim} \frac{C}{d^{(k+1)/2}}d^k= Cd^{(k-1)/2}, \end{equation}
where we used that for $G$ a GUE matrix on $\C^n$, $\E\tr(G^2)\sim_{n\rightarrow+\infty} n^2$ and $\E[\tr(G^2)]^{1/2}\sim_{n\rightarrow+\infty} n$ (see e.g. Ref.~\cite{AGZ}, Chapter 2, for a proof).

Let us now show that the functions $G\mapsto\tr(\rho_GG)$ and $G\mapsto\sup_{\rho\in\cS_{(2)}}\tr(\rho G)$ concentrate around their respective average values. In that aim, we shall make use of the following Gaussian deviation inequality (see e.g. Ref.~\cite{Pisier}, Chapter 2): Assume that $f$ is a function satisfying, for any Gaussian random variables $G,H$, $|f(G)-f(H)|\leq \sigma_{G,H}\|G-H\|_{HS}$ for some $\sigma_{G,H}$ such that $\E\sigma_{G,H}\leq L$. Then, for any $\e>0$,
\begin{align} \P\left( \left|f-\E f\right| > \e \right) \leq e^{-c_0\e^2/L^2}, \end{align}
where $c_0>0$ is a universal constant.

Define $f:G\in GUE(\C^n)\mapsto\tr(G^2)$, and $f_{\Sigma}:G\in GUE(\C^n\otimes\C^n)\mapsto\sup_{\rho\in\Sigma}\tr(\rho G)$, for any given set of states $\Sigma$ on $\C^n\otimes\C^n$. We have first
\begin{align} |f(G)-f(H)| =\, & |\tr(GG^{\dagger})-\tr(HH^{\dagger})|\\
\leq\, & \|GG^{\dagger}-HH^{\dagger}\|_{tr}\\
\leq\, & (\|G\|_{HS}+\|H\|_{HS})\|G-H\|_{HS}, \end{align}
where the last inequality is by the triangle inequality, the Cauchy--Schwarz inequality, and the invariance of $\|\cdot\|_{HS}$ under conjugate transposition, after noticing that $GG^{\dagger}-HH^{\dagger}=G(G^{\dagger}-H^{\dagger})+(G-H)H^{\dagger}$. And second
\begin{align} \left|f_{\Sigma}(G)-f_{\Sigma}(H)\right| =\, & \left|\underset{\rho\in\Sigma}{\sup}\tr(\rho G) - \underset{\rho\in\Sigma}{\sup}\tr(\rho H)\right|\\
\leq\, & \underset{\rho\in\Sigma}{\sup}\left|\tr(\rho[G-H])\right|\\
\leq\, & \|G-H\|_{op}\\
\leq\, & \|G-H\|_{HS}, \end{align}
where the next to last inequality is because $\Sigma$ is a subset of the trace class unit sphere of Hermitians on $\C^n\otimes\C^n$.

Hence, the functions $f$ and $f_{\Sigma}$ satisfy the hypotheses of the Gaussian deviation inequality with $L=2\gamma_{d^k}\sim_{d\rightarrow+\infty}2d^k$ and $L=1$ respectively. So by the mean estimates \eqref{eq:E1} and \eqref{eq:E2}, we have that for any $0<\e<1$,
\begin{equation} \label{eq:F1} \P\left( \tr\left(\rho_GG\right) < (1-\e)\alpha d^{k/2} \right) \leq \exp\left(-c_0\left(\e\alpha d^{2k}\right)^2/\left(2d^k\right)^2\right) = \exp\left(-c'_0\e^2d^{2k}\right), \end{equation}
\begin{equation} \label{eq:F2} \P\left( \underset{\rho\in\cS_{(2)}}{\sup}\tr(\rho G) > (1+\e)Cd^{(k-1)/2} \right) \leq \exp\left(-c_0\left(\e Cd^{(k-1)/2}\right)^2\right) = \exp\left(-c'_0\e^2d^{k-1}\right). \end{equation}

As a consequence, we have that for any $\beta_d$ satisfying $1/2Cd^{(k-1)/2}\leq \beta_d\leq 3/2\alpha d^{k/2}$, the Hermitian $M=\mathbbm{1}-\beta_dG$ on $(\C^d)^{\otimes k}$ is a GME witness for $\rho_G$ with probability greater than $1-\exp(-cd^{k-1})$, where $c>0$ is a universal constant. Indeed, choosing $\e=1/6$ in Eq.~\eqref{eq:F1} and $\e=1/2$ in Eq.~\eqref{eq:F2}, we get
\begin{align} \P\left( \tr\left(\rho_GM\right) > -\frac{1}{4} \right) \leq e^{-cd^{2k}}\ \text{and}\ \P\left( \underset{\rho\in\cS_{(2)}}{\sup}\tr(\rho M) < \frac{1}{4} \right) \leq e^{-cd^{k-1}}, \end{align}
which concludes the proof. \hfill $\square$

\medskip

We may actually say even more on the random state $\rho_G$ defined by Eq.~\eqref{eq:rho_G}. Indeed, define for all $0<\e<1$ the state $\widetilde{\rho}_G(\e)$ on $(\C^d)^{\otimes k}$ by
\begin{align} \widetilde{\rho}_G(\e)= \e\rho_G+(1-\e)\frac{\mathbbm{1}}{d^k} = \frac{1}{d^k}\left(\mathbbm{1}+\e\frac{\alpha}{d^{k/2}}G\right). \end{align}
Then, what the proof of Proposition \ref{prop:GME} additionally tells us is that, as long as $\e\geq \gamma/\sqrt{d}$, for some constant $\gamma>0$, $\widetilde{\rho}_G(\e)$ is with high probability not bi-separable. This means that $\rho_G$ is typically a fully PPT state on $(\C^d)^{\otimes k}$ which is not bi-separable, and whose random robustness of genuinely multipartite entanglement (as defined in Ref.~\cite{VT}) additionally grows at least as $\sqrt{d}$ when $d\rightarrow+\infty$.

\section{Conclusion}

The problem of characterizing genuine multipartite entanglement and biseparability
is difficult. Therefore, a natural approach lies in the relaxation of the definition of
biseparability: Instead of considering states which are separable with respect to some
bipartition, one replaces this set by an appropriate superset, e.g. defined by the PPT
condition or some other positive but not completely positive map.

In this paper we investigated this approach from several perspectives. First, we
established how this relaxation approach with positive maps can be evaluated with semidefinite
programming and how entanglement witnesses can be constructed for this problem. Then, we
showed that, in principle, also other relaxations, besides those obtained from positive maps (e.g. based on
the CCNR criterion), are possible. Finally, we studied the accurateness of the relaxation approach.
We proved rigorous bounds on the volume of the set of biseparable states as well as on the volume of the set of
states which are PPT for any cut. In this way, we showed that in the limit of large dimensional multipartite systems, the relaxation approach detects only a small fraction of the
multiparticle entangled states. It must be stressed, however, that this does not mean
that the relaxation approach is not fruitful: First, it is a well known fact from the theory
of two-particle entanglement that simple entanglement criteria miss most of the states if
the dimension of the local spaces increases \cite{beigi,AS}. Second, from a practical point of view,
the relaxation approaches are clearly the best tools for characterizing multiparticle
entanglement available at the moment \cite{MHRBS, PPTmixer}.

For future research, there are many open questions to address: First, a systematic
approach for the various positive maps besides the transposition would be desirable.
Then, an approach for characterizing separability classes besides biseparability
(e.g. triseparability) would be useful. Finally, methods to certify the Schmidt-rank or the dimensionality of entanglement \cite{MHranks} in high-dimensional systems are needed for current experiments. Investigating the generic scaling of these quantities could also be of interest.

\bigskip
\emph{Acknowledgements.}
We thank T. Moroder and S. W\"olk for insightful discussions. Furthermore we want to thank T. Moroder for providing a MATLAB package for the ``Choi-mixer''. This work has been supported by the
EU (STREP-Project RAQUEL, ERC advanced grant IRQUAT, Marie Curie CIG 293993/ENFOQI),
the FQXi Fund (Silicon Valley Community Foundation),
the Spanish MINECO (project No.~FIS2013-40627-P and
Juan de la Cierva fellowship JCI 2012-14155),
the Generalitat de Catalunya (CIRIT Project No.~2014 SGR 966), the French CNRS (ANR projects OSQPI 11-BS01-0008 and Stoq 14-CE25-0033) and the DFG.

\appendix

\section{Classical convex geometry}
\label{ap:convex-geometry}

We work in the Euclidean space $\R^n$, where we denote by $\|\cdot\|$ the Euclidean norm, induced by the inner product $\langle\cdot,\cdot\rangle$. We denote by $\vol(\cdot)$ the $n$-dimensional Lebesgue measure. A {\em convex body} $K \subset \R^n$ is a convex compact set with non-empty interior. 


If $u$ is a vector from the unit Euclidean sphere $S^{n-1}$, the {\em support
function} of $K$ in the direction $u$ is
\begin{align} h_K(u):=\max_{x\in K}
\langle x, u\rangle. \end{align}
Note that $h_K(u)$ is the distance from the origin to the hyperplane tangent to $K$ in the direction $u$.

Two global invariants associated to a convex body $K \subset \R^n$, the {\em volume radius} and the {\em mean width}, play an important role in our proofs.

\begin{definition} \label{def:vrad}
The {\em volume radius} of a convex body $K \subset \R^n$ is defined as, denoting by $B^n$ the unit Euclidean ball of $\R^n$,
\begin{align} \vrad (K) := \left( \frac{\vol K}{\vol B^n} \right)^{1/n}.\end{align}
In words, $\vrad(K)$ is the radius of the Euclidean ball with same volume as $K$.
\end{definition}

\begin{definition} \label{def:w}
The {\em mean width} of a subset $K \subset \R^n$ is defined as
\begin{align} w (K):=\int _{S^{n-1}} \max_{x\in K} \langle x,u \rangle \,\mathrm{d}\sigma(u) ,\end{align}
where $\mathrm{d}\sigma(u)$ is the normalized spherical measure on the unit Euclidean sphere $S^{n-1}$ of $\R^n$.
If $K$ is a convex body, we have
\begin{align} w (K):=\int _{S^{n-1}}h_K(u)\,\mathrm{d}\sigma(u).\end{align}
\end{definition}

The inequality below (see e.g. Ref.~\cite{Pisier}, Corollary 1.4) is a fundamental result which compares the volume radius and the mean
width.

\begin{theorem}[Urysohn inequality]
\label{th:urysohn}
For any convex body $K \subset \R^n$, we have
\begin{align} \vrad(K) \leq w(K) .\end{align}
\end{theorem}

It is convenient to compute the mean width using Gaussian rather than spherical integration. Let $G$ be a standard Gaussian vector in $\R^n$, i.e. such that its coordinates, in any orthonormal basis, are independent and following a Gaussian distribution with mean $0$ and variance $1$. Setting $\gamma_n = \E \|G\|\sim \sqrt{n}$, we have, for any compact set $K \subset \R^n$,
\begin{align} w_G(K) := \E \left(\max_{x \in K} \langle G,x \rangle\right) = \gamma_n w(K) .\end{align}

We also need the two following lemmas, which are incarnations of the familiar ``union bound''. They appear for example in Ref.~\cite{LT}, Chapter 3, under the equivalent formulation via suprema of Gaussian processes.

\begin{lemma}[Bounding the mean width of a polytope] \label{lemma:mean-width-polytope}
Let $v_1,\ldots,v_N$ be points in $\R^n$ such that $v_i\in\lambda B^n$ for every index $1\leq i\leq N$ (where $B^n$ denotes the unit Euclidean ball of $\R^n$). Then
\begin{align} w\left(\conv(v_1,\ldots,v_N)\right) \leq \lambda \sqrt{\frac{2\ln N}{n}}. \end{align}
\end{lemma}

\begin{lemma}[Bounding the mean width of a union] \label{lemma:mean-width-union-bound}
Let $K_1,\dots,K_N$ be convex sets in $\R^n$ such that $K_i \subset \lambda B^n$ for every index $1\leq i\leq N$ (where $B^n$ denotes the unit Euclidean ball of $\R^n$). Then
\begin{align} w \left( \conv \left( \bigcup_{i=1}^N K_i \right) \right) \leq 2 \left(\max_{1 \leq i \leq N} w(K_i) +  \lambda \sqrt{\frac{2\ln N}{n}} \right).\end{align}
\end{lemma}

Finally, we use repeatedly the following result, established in Ref.~\cite{MP}, Corollary 3.

\begin{theorem}[Milman--Pajor inequality]
\label{th:Milman-Pajor}
Let $K,L$ be convex bodies in $\R^n$ with the same center of mass. Then
\begin{align} \mathrm{vrad}(K\cap L)\mathrm{vrad}(K-L)\geq\mathrm{vrad}(K)\mathrm{vrad}(L),\end{align}
where $K-L=\left\{x-y \st x\in K,\ y\in L\right\}$ stands for the Minkowski sum of the convex bodies $K$ and $-L$.
\end{theorem}

Choosing $K=-L$ in Theorem \ref{th:Milman-Pajor} yields the following corollary.

\begin{corollary}
\label{corollary:Milman-Pajor}
If $K$ is a convex body in $\R^n$ with center of mass at the origin, then
\begin{align}\mathrm{vrad}(K\cap -K)\geq\frac{1}{2}\mathrm{vrad}(K),\end{align}
and more generally for any orthogonal transformation $\theta$,
\begin{align}\mathrm{vrad}(K\cap\theta(K))\geq\frac{1}{2}\frac{\mathrm{vrad}(K)^2}{w(K)}.\end{align}
\end{corollary}

The latter inequality is simply because, on the one hand, $\mathrm{vrad}(\theta(K))=\mathrm{vrad}(K)$, and on the other hand, by Theorem \ref{th:urysohn}, $\mathrm{vrad}(K-\theta(K))\leq w(K-\theta(K)) \leq w(K)+w(\theta(K)) = 2w(K)$.

We typically use Corollary \ref{corollary:Milman-Pajor} in the following way: if $K$ is a convex body with center of
mass at the origin which satisfies a ``reverse'' Urysohn inequality,
i.e. $\mathrm{vrad}(K) \geq \alpha w(K)$ for some constant $\alpha$, we conclude that the volume radius of $K \cap \theta(K)$ is comparable to the volume radius of $K$.




\end{document}